\documentclass[conference]{IEEEtran}
\IEEEoverridecommandlockouts

\usepackage{cite}
\usepackage{amsmath,amssymb,bm}
\usepackage{cleveref}
\usepackage{graphicx}
\usepackage{subcaption} 
\usepackage{url}
\usepackage{xcolor}
\usepackage{tabularx}
\usepackage{array}

\DeclareMathOperator*{\argmax}{argmax}
\def\BibTeX{{\rm B\kern-.05em{\sc i\kern-.025em b}\kern-.08em
    T\kern-.1667em\lower.7ex\hbox{E}\kern-.125emX}}

\title{Pricing Innovation Under Latency Constraints: A Mean-Field Analysis of Coded Payload Delivery\thanks{Funding by Optimum}}

\author{\IEEEauthorblockN{1\textsuperscript{st} Muriel M\'edard}
\IEEEauthorblockA{\textit{Optimum}\\ Cambridge, MA, USA\\ mmedard@getoptimum.xyz}
\and
\IEEEauthorblockN{2\textsuperscript{nd} Tarun Chitra}
\IEEEauthorblockA{\textit{Gauntlet}\\ New York, NY, USA\\ tarun@gauntlet.network}
\and
\IEEEauthorblockN{3\textsuperscript{rd} Moritz Grundei}
\IEEEauthorblockA{\textit{Optimum}\\ Munich, Germany\\ moritz@getoptimum.xyz}
\and
\IEEEauthorblockN{4\textsuperscript{th} Sajida Zouarhi}
\IEEEauthorblockA{\textit{Optimum}\\ Varna, Bulgaria\\ sajida@getoptimum.xyz}
}

\begin{document}
\maketitle

\begin{abstract}
We study pricing mechanisms for low-latency payload delivery in settings where participant rewards depend on the time required to reconstruct a payload. In such environments, the decoding-time distribution determines deadline-meeting probabilities and therefore bounds a participant’s willingness to pay for additional delivery rate. Using a mean-field formulation, we derive price–rate bounds from simple stochastic arrival models and instantiate them for (i) unsharded transmission and (ii) sharded delivery under three regimes: uncoded sharding, fixed-rate erasure coding, and rateless coding. These bounds yield a comparative characterization of how symbol usefulness translates into economic value under deadline-driven utilities.

We further analyze a two-lane service consisting of a base lane and an Random Linear Network Coding (RLNC) fast lane. In this Turbo decoding setting, a receiver combines shards arriving via both lanes to minimize time-to-decode. Under a fixed base-lane price–rate pair and an aggregate rate constraint, we derive a fast-lane pricing bound and show how even modest additional RLNC rate can generate measurable utility gains, depending on the base-lane propagation regime. The framework extends naturally to stepwise reward schedules with multiple deadlines, and we illustrate its applicability on representative scenarios motivated by blockchain message dissemination and latency-sensitive competition.
\end{abstract}


\section{Introduction}

Low-latency data delivery has economic value whenever the usefulness of information depends on when it is received. In distributed systems with hard timing constraints, the key performance object is therefore not only throughput or average delay, but the distribution of end-to-end completion times. Small changes in the right tail can translate into large, discontinuous changes in payoff when deadlines are missed. Contemporary blockchain networks are a canonical example: core messages such as blocks and attestations, and increasingly blob-sized data payloads, must be disseminated under protocol-defined clocks over a gossip-based peer-to-peer network (e.g., GossipSub). In both cooperative settings (consensus rewards and penalties) and adversarial settings (latency-sensitive competition), participants benefit from reconstructing and reacting to a a disseminated payload as quickly as possible.

Ethereum makes this deadline sensitivity explicit. Consensus proceeds in fixed-length slots, and validators must receive the head block and broadcast an attestation within a narrow time window. When block propagation is slow, fewer validators can attest in time, and even correct attestations earn less if they are included later: payouts are discounted by inclusion delay, so expected staking yield depends on the distribution of propagation and aggregation latency rather than on average delay alone \cite{neroeth2024attestations_block_propagation_timing_games,neroeth2024deep_diving_attestations,oz2023time,schwarzschilling2023timing}. This latency dependence extends beyond becon blocks. Proto-danksharding (EIP-4844) introduced blob-carrying transactions, moving the dissemination of large data payloads into the consensus critical path \cite{eip4844,soltani2024eip4844delay}. With PeerDAS (EIP-7594), validators are expected to participate in data availability sampling and reconstruction of erasure-coded blob data, so the relevant deadline is no longer just receiving a block, but completing sufficient blob shard downloads to confidently attest \cite{eip7594}.

Latency also has direct competitive value. Under proposer--builder separation (PBS) as enabled through MEV-Boost, each slot additionally hosts a real-time block-building auction \cite{daian2019flash}: builders iteratively submit candidate blocks and bids via relays, and the proposer must obtain and publish the winning payload early enough for the network to process it. Delays at any link of this pipeline have first-order economic effects \cite{schwarzschilling2023timing}. Searchers continuously monitor latency-critical public signals, e.g., mempool transactions, DEX state updates, oracle updates, and cross-venue price moves, and must decode these signals, compute an optimal bundle, and deliver it to a builder before the builder’s internal cutoff for simulation and inclusion. In competitive MEV settings, even sub-second differences can decide whether an arbitrage, liquidation, or sandwich opportunity is captured or lost \cite{flashbots2025infrastructurerace}. Builders compete in an open, rapidly-updating auction with an end-of-slot deadline; the strategic value of information rises over the slot because late-arriving order flow and last-moment state changes can dominate block value \cite{wahrstatter2023time,schwarzschilling2023timing}. Empirically and theoretically, this induces “timing games’’ and strong incentives to reduce network latency to relays and to the proposer \cite{yang2024buildermarket,schwarzschilling2023timing,oz2023time,pai2023integrated}. Operationally, relays also enforce safety cutoffs on late payload delivery, so a proposer that requests the winning payload too late risks a missed proposal and a discrete loss in rewards. These mechanics convert propagation and decoding delay directly into expected revenue, and they create structural incentives for geographic and infrastructural centralization in the block-building supply chain \cite{yang2024buildermarket,pai2023integrated}.

A central observation of this paper is that different coding schemes can differ substantially in these decoding-time statistics even at comparable bandwidth, because coding changes what it means for a received unit of data to be useful. Under uncoded or fixed-rate coded payload delivery, received shards may be redundant since they might be a duplicate of a fragment already received; with rateless coding in general and Random linear network coding (RLNC) in particular, newly received symbols are innovative with high probability until decoding is achieved \cite{ho2006random}, \cite{koetter2003algebraic}. These distinctions affect the probability of meeting deadlines and therefore translate directly into different willingness-to-pay for additional delivery rate for nodes that derive economic value from timely payload reception. Our goal is to make this link explicit in a minimal model that is suitable for comparative economic insight.

This paper is structured as follows. Section \ref{sec:notation} introduces the notation used throughout. Section \ref{sec:system_model} analyzes the arrival-time statistics of different sharding and coding schemes under a simplified Poisson shard-arrival model. In Section \ref{sec:mean_field_price_bound}, we define the economic utility function as a mean-field utility and derive a price bound for the economic value of decoding time. Section \ref{sec:turbo_pricing} specializes this analysis to derive the pricing bound for Turbo-coded shards. We conclude with two stylized examples, motivated by validator utility in Ethereum consensus and searcher competition in an MEV-style auction, modeled as a \(k\)-first race.

\section{Notation}
\label{sec:notation}
We denote scalars by lowercase letters and random variables by uppercase letters (e.g., $X$). The cumulative distribution function (CDF) of $X$ is written as $F_X(x; k)$, where $x$ is the realized value and $k$ denotes a distribution parameters. We write $E[X]$ for the expectation of $X$. Stochastic counting processes (e.g., Poisson processes) are denoted by $S(t)$. Superscripts $(1)$ and $(2)$ indicate whether a quantity belongs to the base lane or the fast lane, respectively. For example, $S^{(1)}(t)$ and $\lambda^{(1)}$ denote the counting process and arrival rate of shards in the base lane.

\begin{figure*}[t]
    \centering
    \includegraphics[width=\linewidth]{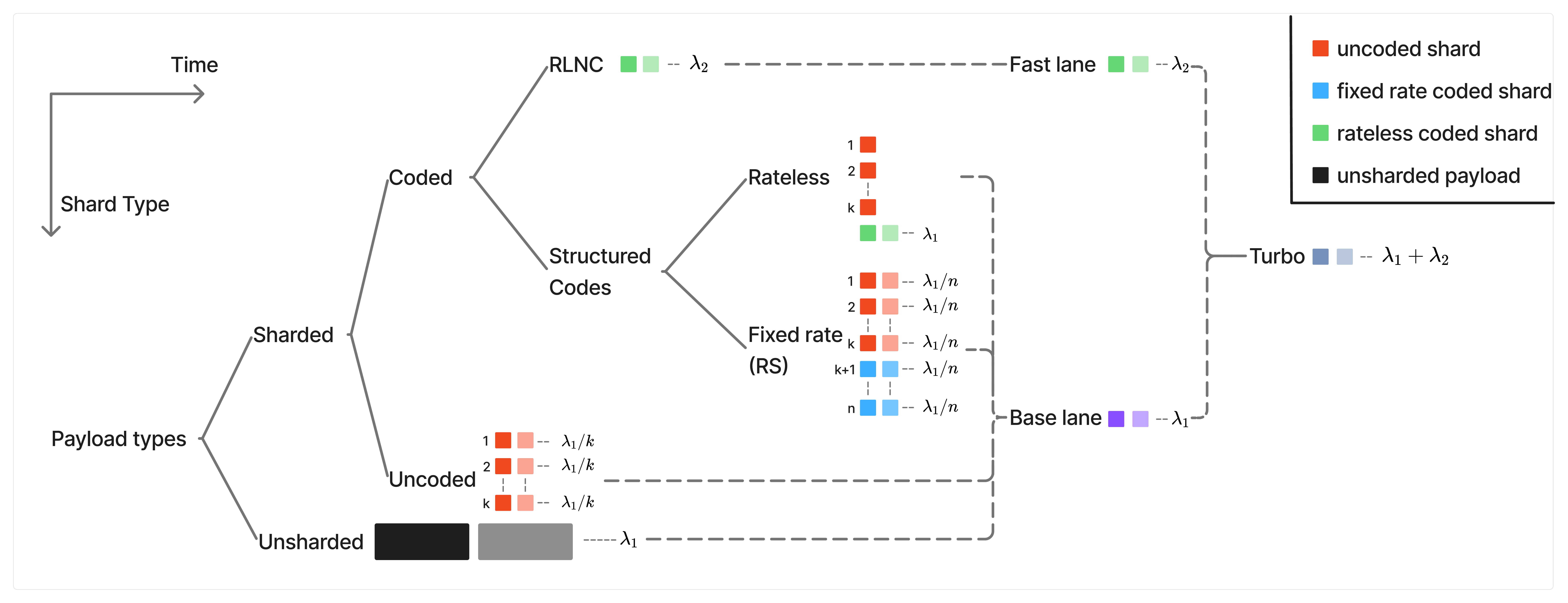}
    \caption{Turbo approach for message propagation. Turbo shards consist of one type of several possible base lane shard types as well as RLNC fast lane shards.}
    \label{fig:turbo_structure}
\end{figure*}

\section{System Model}
\label{sec:system_model}
Information transmitted a network is segmented into packets, which are then propagated as discrete units across the underlying infrastructure. From the receiver's perspective, packet arrival times are shaped by a range of stochastic effects, including queueing delays, packet loss, retransmissions, route changes, and other forms of time-varying network behavior. It is therefore natural to model packet arrival times as random variables. As a first approximation to this arrival process, we adopt a Poisson model. This is not intended as a protocol-faithful description of packet propagation, but rather as a standard tractable baseline for comparative analysis. Poisson arrival processes have long played this role in the queueing and networking literature, including in classical treatments such as Kleinrock's Queueing Systems ~\cite[Chapter~2.2]{kleinrock1975queueing} and Bertsekas and Gallager's Data Networks ~\cite[Chapter~3]{bertsekas1992data}, where they serve as an analytically convenient first approximation to packet traffic. We use the Poisson model here as a reference point for understanding how different delivery and coding choices affect decoding latency. The associated independence assumption, in line with the Palm--Khintchine approximation, intentionally abstracts away correlations induced by topology, retransmissions, batching, and shared bottlenecks. This enables us to compare decoding-time distributions under a common baseline model and to attribute differences primarily to the delivery regime itself rather than to secondary features of a more detailed propagation model.

We now describe several different message types and their stochastic properties within the limits of this model, as illustrated in Fig.~\ref{fig:turbo_structure}.
We consider two dissemination paradigms: unsharded data and sharded data. For sharded data, we assume that the arrival process is evenly divided across shards. Specifically, if a message is divided into $k$ shards, we model $k$ parallel Poisson arrival streams at the receiver, each with rate $\lambda/k$. For an unsharded payload, we model a single arrival process with rate $\lambda/k$ to account for the $k$-times larger message size to account for the bandwidth constraint.

\subsection{Delay Models}
\label{sec:delay_models}

For unsharded data, the arrival time $X_{\text{unsharded}}$ follows an exponential distribution with rate $\lambda/k$, that is $X_{\text{unsharded}} \sim \text{Exp}(\lambda/k)$.

For the sharded, uncoded payload, the arrival time $X_{\text{uncoded}}$ is determined by the slowest shard. The cumulative distribution function (CDF) of $X_{\text{uncoded}}$ is therefore

\begin{equation}
    \begin{aligned}
        F_{X_\text{uncoded}}(\tau; k, \lambda)
            &= P(X_\text{uncoded} \leq \tau) \\
            &= (1 - e^{-\tau \lambda/k})^{k}.
    \end{aligned}
    \label{eq:uncoded_cdf}
\end{equation}

For rateless codes such as RLNC \cite{koetter2003algebraic} or Raptor codes \cite{shokrollahi2006raptor}, shards are not individualized. We therefore model a single arrival process with rate $\lambda$, and decoding requires the reception of $k$ coded shards. The decoding time $X_{\text{rateless}}$ is the sum of $k$ exponential random variables with rate $\lambda$, which follows an Erlang distribution. Its cumulative distribution function is

\begin{equation}
    F_{X_{\text{rateless}}}(x; k, \lambda)
        = 1 - \sum_{i=0}^{k-1} \frac{(\lambda x)^{i} e^{-\lambda x}}{i!}.
    \label{eq:rateless_cdf}
\end{equation}

Another approach to sharding data is to use a fixed-rate $(n, k)$ code, such as a Reed–Solomon (RS) code. In this case, we grant each shard an equal probability of arrival even though arrival orders typically exhibit noticeable correlation. We model the timely-decoding probability by treating the arrivals of the $n$ coded shards as statistically independent and focusing on the event that at least $k$ distinct coded shards have arrived by time $\tau$. For notational simplicity, we refer to RS codes as the representative fixed-rate code, although the results apply to any fixed-rate MDS code.

\begin{align}
    F_{X_{\text{RS}}}(\tau; k, n, \lambda)
    = 1 - \sum_{j=0}^{k-1} {n \choose j}
        \left( 1 - e^{-\frac{\lambda \tau}{n}} \right)^{j}
        e^{-\frac{(n-j)\lambda \tau}{n}}.
    \label{eq:RS_cdf}
\end{align}

We note that for a fixed $k$ but $n\rightarrow\infty$, the success probability $p_n$ in the binomial tail becomes
$p_n=(1-e^{-\lambda\tau/n}) \rightarrow\lambda\tau/n$ in the first order approximation. This means that the Poisson limit theorem applies here as we have $np_n = \lambda\tau$ as $n\rightarrow\infty$ from which follows that the probability of $k$ distinct shards arriving follows an Erlang distribution itself. For the probability that at least $k$ shards arrive, this means that

\begin{align*}
    F_{X_{RS}}(\tau; n\rightarrow\infty, k, \lambda) = 1- \sum_{j=0}^{k-1}\frac{(\lambda\tau)^je^{-\lambda\tau}}{j!},
\end{align*}

which corresponds to the CDF of the Erlang distribution that characterizes the timely arrival probability of $k$ rateless coded shards. This result can be interpreted as follows: as the coding rate decreases by allowing the number of coded shards $n$ to grow, the likelihood of receiving a non-innovative shard becomes negligible.

There is one important caveat when considering fixed rate codes in general and RS codes in particular \cite{reed1960polynomial}. For example, an RS code requires $n \le |\mathbb{F}|$ because it is defined by evaluating a polynomial at $n$ distinct field elements (locators). In practice, RS codes deployed in systems such as Celestia or Solana’s Rotor are typically defined over extension fields with 2-byte elements, giving $|\mathbb{F}| = 2^{16}$ and therefore $n \le 2^{16}$.

\begin{figure}
    \centering
    \includegraphics[width=\linewidth]{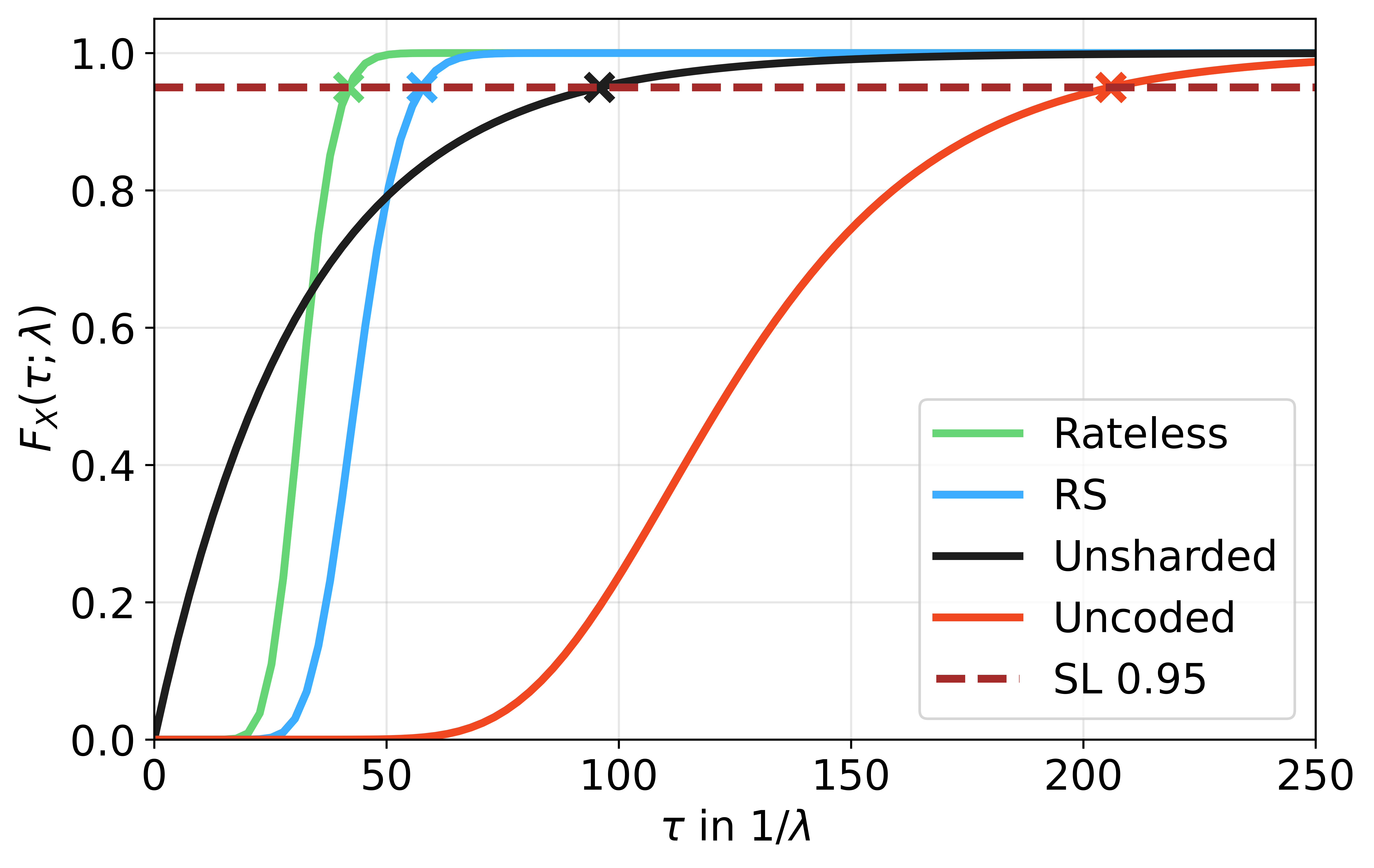}
    \caption{Comparison of cumulative distribution functions of arrival times when considering unsharded payloads and sharded payloads. For sharding, we set $k=32$ and $n=64$ (only relevant for fixed-rate coded payloads). Further, an exemplary service level (SL) of 95\% arrival probability is indicated.}
    \label{fig:cdfs}
\end{figure}

The CDFs in \cref{eq:uncoded_cdf,eq:rateless_cdf,eq:RS_cdf} are compared in Fig.~\ref{fig:cdfs} for a sharding factor $k=32$ and $n=64$ coded shards for fixed rate coded payloads as used for example in Solana's Rotor \cite{kniep2025solana}. Among the sharded payload schemes, and for a fixed arrival rate, rateless codes achieve a higher probability of timely arrival for all $\tau$ compared with uncoded or fixed-rate coded shards. The unsharded payload, on the other hand, exhibits a substantially higher probability of early arrival for small values of $\tau$ relative to the sharded cases. However, if a node requires a high probability of successful reception for example, a service level (SL) of 95\%, then the required waiting time becomes significantly longer for the unsharded payload. In our examples, this waiting time is roughly twice as long as for rateless-coded payloads and about 1.7 times as long as for RS-coded payloads.

The intuitive explanation is that coding effectively averages arrival times across many shards, which reduces the chance of early completion but also diminishes the probability of long delays. Conversely, unsharded payloads have a non negligible probability of very fast arrivals but have heavier tails. The same reasoning applies to the rate $\lambda$, since in all CDF expressions in \cref{eq:rateless_cdf,eq:uncoded_cdf,eq:RS_cdf}, the rate appears only through the product $\lambda \tau$.

\subsection{Turbo Delay Model}
\label{sec:turbo}

Now, for the Turbo setup as displayed in Fig. \ref{fig:turbo_structure}, we have to consider shards / payloads from two different sources: 1) The base lane which might be unsharded, sharded and uncoded or sharded and rateless or fixed-rate coded. We introduce the stochastic process $S^{(1)}(\tau)$ which counts the number of innovative shards that have arrived by time $\tau$. 2) The fast lane which uses RLNC coded shards. The number of innovative shards $S^{(2)}(\tau)$ is Poisson distributed with rate $\lambda^{(2)}$ since we assume the probability of linearly dependent shards to be negligibly small \cite{networkCodingForEngineers}.

To model the time until decoding in the turbo scenario $X_T$, we must consider two different scenarios. First, we consider the scenario where the base lane is sharded. In this case, we can make use of the unique property of RLNC to use any coded or uncoded shard in its decoding process since all these are just degenerate forms of random coding as embodied by RLNC \cite{neu2019babel, hellge2016multi}. The number of unique shards from the base lane and the fast lane must sum up at least to the sharding factor $k$:

\begin{align*}
    F_{X_T}(\tau; \lambda^{(1)}, \lambda^{(2)}, k) &= P(\{S^{(1)}(\tau) + S^{(2)}(\tau)\ge k\}) \\
    & =  1 - P(\{S^{(1)}(\tau) + S^{(2)}(\tau)\le k-1\})\\
    & = 1 - \sum_{i=0}^{k-1}P(\{S^{(1)}(\tau) + S^{(2)}(\tau) = i\})\\
    & = 1 - \sum_{i=0}^{k-1}\sum_{j=0}^{i}P(\{S^{(1)}(\tau) = j\})\\&\;\;\; P(\{S^{(2)}(\tau) = i-j\})
\end{align*}

where the last line follows from the independence of $S^{(1)}$ and $S^{(2)}$. Table \ref{tab:unique_shard_process} compares the distribution of the innovative shard counting process. If the base lane happens to employ rateless coding as well, then the $X_T$ is distributed like $X_{rateless}$ (Erlang with rate $\lambda_1+\lambda_2$)

If the base lane is unsharded, the decoding time $X_T$ comes down to a race between the base lane and the fast lane

\begin{align*}
    F_{X_T}(\tau; \lambda^{(1)}, \lambda^{(2)}, k) 
    & = P(\{\min(X_{\text{unsharded}}, X_{\text{rateless}}) \leq \tau\})\\
    & = 1 - (1-F_{X_{\text{unsharded}}}(\tau; k, \lambda^{(1)}))\\&\;\;\;(1-F_{X_{\text{rateless}}}(\tau; k, \lambda^{(2)}))
\end{align*}

\begin{table}[ht]
\renewcommand{\arraystretch}{1.2}
\begin{tabular}{ll}
\hline
\textbf{Shard type} & \textbf{Distribution of $S(\tau)$} \\
\hline
Uncoded & $\text{Binomial}(k, 1-e^{-\lambda\tau/k})$ \\
Fixed-rate coded & $\text{Binomial}(n, 1-e^{-\lambda\tau/n})$ \\
Rateless coded & $\text{Poisson}(\lambda\tau)$ \\
\end{tabular}
\caption{Distribution of the shard counting process $S(\tau)$ under different coding schemes.}
\label{tab:unique_shard_process}
\end{table}

\section{Mean Field Price Bound}
\label{sec:mean_field_price_bound}
In blockchain systems, certain groups of nodes derive utility from receiving a message early. 
We capture this by a random variable $U$, which depends on the message arrival (or decoding) time $X$ at a node.

From the perspective of an individual node, the central question is: what price $p$ is the node willing to pay per shard, given that shards arrive at rate $\lambda$? 
The total expenditure per unit time (e.g., per slot) is therefore $\lambda p$.

A mean-field model provides an upper bound on a node's willingness to pay. In this framework, the willingness to pay cannot exceed the expected utility obtained from receiving shards, i.e.,

\begin{equation}
    p\lambda \leq \mathbb{E}\!\left[U(X)\right].
    \label{eq:Utility}
\end{equation}

\begin{figure}[t]
    \centering
    \includegraphics[width=\linewidth]{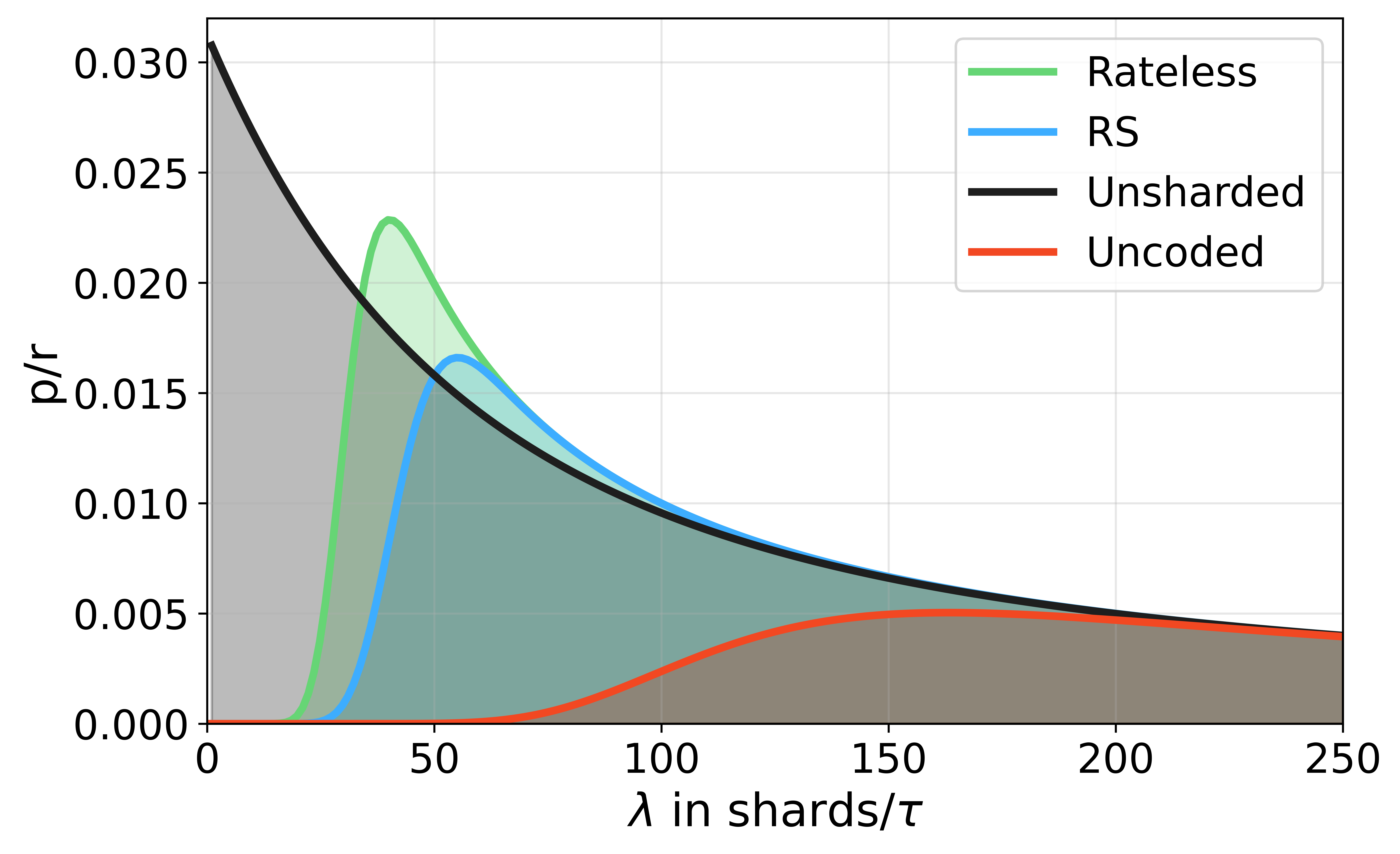}
    \caption{Comparison between rateless coding, fixed-rate coding, and uncoded transmission (base lane proxy). We set $n=64$ and $k=32$. Left: Distributional behavior of arrival time; Right: price bound in mean-field model for single delay, single reward setup.}
    \label{fig:comparison_price_rate_bounds}
\end{figure}

Any pair $(p,\lambda)$ satisfying \eqref{eq:Utility} constitutes a feasible price--service pair for the node.
Figure~\ref{fig:comparison_price_rate_bounds} compares an exemplary instantiation of these bounds. 
The price-per-rate bound for rateless shards is strictly higher than that for fixed-rate or uncoded shards: individual rateless shards provide greater marginal value because they are, with high probability, innovative (i.e., linearly independent and thus uniquely useful for decoding)~\cite{networkCodingForEngineers}.

\section{Turbo Pricing Model}\label{sec:turbo_pricing}

\begin{figure*}[t]
    \centering
    \begin{subfigure}{0.49\textwidth}
        \centering
        \includegraphics[width=\linewidth]{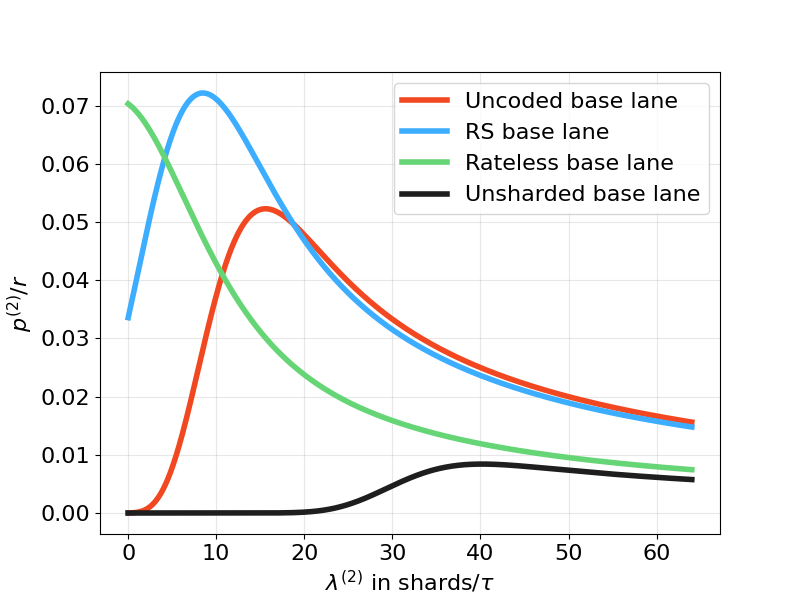}
    \end{subfigure}\hfill
    \begin{subfigure}{0.49\textwidth}
        \centering
        \includegraphics[width=\linewidth]{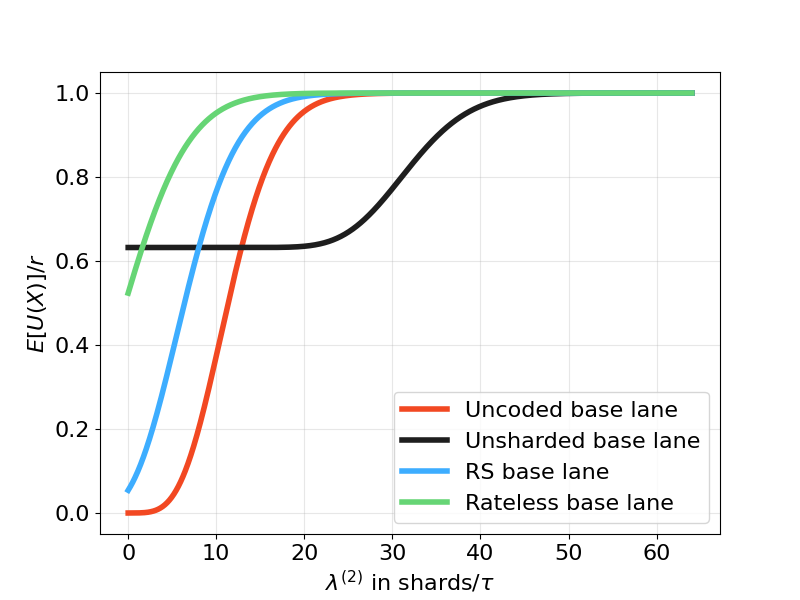}
    \end{subfigure}
    \caption{Price and corresponding node revenue per fast-lane user, relative to the reward $r$ obtained when decoding completes by time $\tau$, using Turbo decoding with unsharded, uncoded, fixed-rate, and rateless-coded base lanes and an RLNC fast lane. The base-lane rate is $\lambda^{(1)} = 32\,\text{shards}/\tau$. We set $k = 32$ and $n = 64$.}
    \label{fig:turbo_price}
\end{figure*}

Since structured coding schemes are constrained instantiations of randomized coding as performed by RLNC, Turbo nodes can make use of all types of shards. This mechanism has been studied in the context of combining different coding schemes for distributed coded storage \cite{neu2019babel, hellge2016multi}.

In the Turbo-node setting, a node may use shards originating from both the base lane and the fast lane, where the latter are explicitly generated for RLNC in a rateless fashion. The base lane supplies unsharded, uncoded, rateless, or fixed-rate coded shards at rate $\lambda^{(1)}$, with arrival time $X^{(1)}$. The fast lane provides RLNC-coded shards at rate $\lambda^{(2)}$, which supplement the base-lane payloads and enable a decoding time $X^{(2)}$. We assume that the base-lane price–rate pair is fixed and that the combined rate of both lanes is bounded above by $\lambda_{\max}$, reflecting nodes' bandwidth constraints.

In principle, the fast lane may operate at a lower rate than the base lane while still achieving an earlier decoding time, because it can exploit all information arriving through both lanes. Hence, we have $X^{(2)} < X^{(1)}$.

For illustration purposes, we consider a single delay–reward pair $(\tau, r)$. The node receives the reward if decoding is completed before time $\tau$, i.e.

\begin{align}
    E[U(X)] = E[r\,\mathbb{I}(X \leq \tau)] = r F_X(\tau).
\end{align}

A node’s willingness to pay for additional fast-lane rate $\lambda^{(2)}$ depends on the incremental reward it expects to obtain from using both base-lane and fast-lane shards. This leads to the condition

\begin{equation*}
    p^{(2)} \lambda^{(2)} \leq 
    r F_{X^{(2)}}(\tau; \lambda^{(1)}, \lambda^{(2)})
    - r F_{X^{(1)}}(\tau; \lambda^{(1)}).
\end{equation*}

The objective is to maximize the revenue generated from the fast lane,

\begin{equation*}
    (\lambda^{(2)*}, p^{(2)*})
    = \argmax_{\lambda^{(2)},\, p^{(2)}} \left( p^{(2)} \lambda^{(2)} \right),
\end{equation*}

subject to the constraints

\begin{equation*}
    \begin{aligned}
        &p^{(2)} \leq
        \frac{
            r F_{X^{(2)}}(\tau;\lambda^{(1)}, \lambda^{(2)})
            -
            r F_{X^{(1)}}(\tau;\lambda^{(1)})
        }{\lambda^{(2)}},\\[6pt]
        &\lambda^{(1)} + \lambda^{(2)} \leq \lambda_{\max}.
    \end{aligned}
\end{equation*}

When the price upper bound is positive and revenue increases monotonically with $\lambda^{(2)}$ over the feasible range, the revenue-maximizing solution is obtained by setting $\lambda^{(2)}$ to its maximum value and pricing at the corresponding upper bound. Thus,

\begin{equation*}
    \begin{aligned}
        p^{(2)} &=
            \frac{
                r F_{X^{(2)}}(\tau;\lambda^{(1)}, \lambda_{\max} - \lambda^{(1)})
                -
                r F_{X^{(1)}}(\tau;\lambda^{(1)})
            }{\lambda_{\max} - \lambda^{(1)}},\\[6pt]
        \lambda^{(2)} &= \lambda_{\max} - \lambda^{(1)}.
    \end{aligned}
\end{equation*}

\subsection{Numerical Illustration}

Figure~\ref{fig:turbo_price} shows an example of the resulting price and corresponding node utility in a Turbo configuration, comparing setups with unsharded, uncoded, fixed-rate, and rateless-coded base-lane models as a function of the additional fast-lane rate $\lambda^{(2)}$. We again set \(k = 32\), \(n = 64\), and \(\lambda^{(1)} = 32\,\text{shards}/\tau\), so that the expected number of shards delivered by the base lane within the time interval \(\tau\) is \(k\).. The following observations can be made:

When the fast-lane rate is small, the price a node is willing to pay in the unsharded baseline is also small. This is expected: the synergy between an unsharded payload and RLNC-coded shards is limited, since turbo decoding effectively becomes a race between the base lane and the fast lane—one that the base lane is very likely to win when $\lambda^{(2)}$ is small. Consequently, the node revenue stays mostly unchanged for unsharded base lane payloads at low fast-lane rates. Among the sharded payloads, the uncoded case yields the lowest price because uncoded shards generate the most duplicates, which keeps the probability of timely arrival low even when supplemented with fast-lane shards. In contrast, rateless and RS-coded shards can be used more efficiently by the RLNC decoder: duplicate probability in the base lane is low (or even zero in the rateless case under this model), allowing the decoder to collect $k$ innovative shards more reliably and thus decode the payload sooner.

For higher fast-lane rates ($\lambda^{(2)} \geq 30\;\text{shards}/\tau$), the price charged for the fast lane also increases under an unsharded base lane, since the probability that the fast lane wins the decoding race becomes significant only at such higher rates (see Fig.~\ref{fig:cdfs} for a comparison of the unsharded and rateless-coded CDFs). Among the sharded base-lane payloads, the turbo setup with uncoded shards commands the highest prices, as the fast-lane shards have the strongest marginal effect on the decoding probability. This is reflected in the fact that the fast lane shards enable the node to extract the reward $r$ with high probability. The next-highest price occurs when the base lane uses fixed-rate coded shards, since the fast lane accounts for up to 95\% of the expected revenue in this case as shown in high regions of the revenue plot (Fig. \ref{fig:turbo_price} right).

\section{Examples}\label{sec:examples}
In this section we give two stylized examples of real-world applications that illustrate the applicability of the model and show the potential benefit of the fast lane to the node's expected utility.

\subsection{Multi-Deadline Rewards}
\begin{figure}
    \centering
    \includegraphics[width=\linewidth]{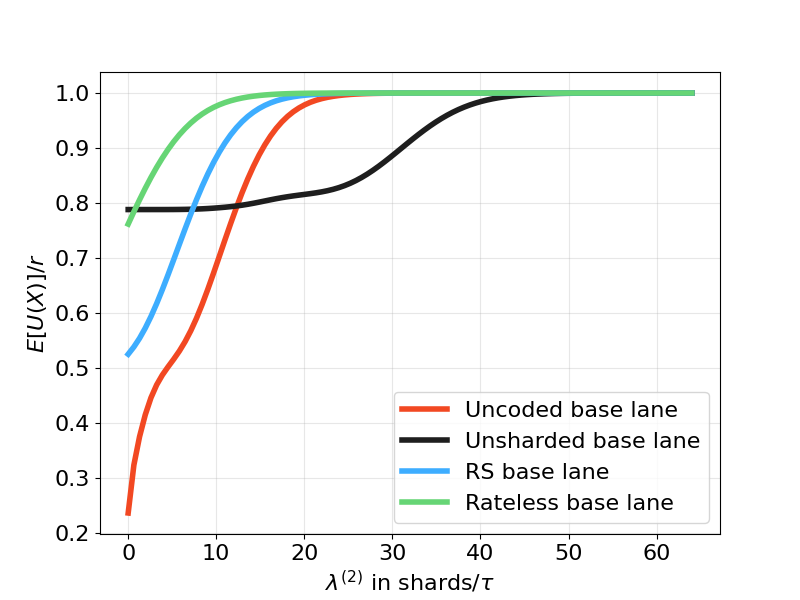}
    \caption{Expected node utility in a multi-deadline multi-reward setup using Turbo decoding with unsharded, uncoded, fixed-rate, and rateless-coded base lanes and an RLNC fast lane. The base-lane rate is $\lambda^{(1)} = 32\,\text{shards}/\tau$. We set $k = 32$ and $n = 64$.}
    \label{fig:expected_utility_multi_deadline}
\end{figure}

First, we study the generalization to a multi-deadline multi-reward setup similar to the attestation reward structure as exhibited in the Ethereum consensus layer.
We model a base delay $\tau$ which would loosely correspond to a slot time and a maximum achievable reward $r$ that is attained if a block is proposed and attested to within the next slot. The reward function is therefore modeled as

\begin{align}
U(X) =
\begin{cases}
r,      & X \le \tau, \\
r/i, & X \in \big((i-1)\tau,\, i\tau\big], \quad i \ge 2.
\end{cases}
\end{align}

The corresponding expected utility $E[U(X)]$ is computed as 
\begin{align*}
    \mathbb{E}[U(X)] = rF_X(\tau_0) + \sum_{i}\frac{r}{i}(F_X(\tau_i) - F_X(\tau_{i-1}))
\end{align*}

An exemplary illustration of the expected utility from additional fast lane shards is displayed in Fig. \ref{fig:expected_utility_multi_deadline}. Similar to the single deadline reward model, the RLNC shards are especially helpful for sharded payloads, even in low fast lane shard regions.

\subsection{Top-k Race}
In MEV, specialized users called searchers bundle user transactions with their own transactions and submit a payload called a bundle.
The relative arrival time of a bundle, or its order statistics, are generally correlated to the magnitude of the reward won.

We define the random variable $X$ as the time when a bundle is received by the builder. $X$ is composed of several delays, however we assume the delay induced by transaction gossiping to be the dominant factor.
For the $i$th  competitor, we define $X_i$ as the arrival time of their bundle at the builder.
We assume that there is a common empirical CDF for $N$ competitors, $\bar F(t) = P(\exists \;i: X_i \leq t) = F_X(t;\lambda)$ for a common $\lambda$.
We assume the searchers are symmetric in that they have the same utility function for their bundles~(e.g.~net profit).
If a searcher is among the first $s$ to arrive, it earns a rank-dependent payoff $r_1 \ge \cdots \ge r_m>0$.
Otherwise, the user's bundle reverts and they have to pay a gas fee of $g>0$.
We therefore use the term Top-k race generically, but denote the number of rewarded participants by $s$ to avoid collision with the shard parameter $k$.
We note that this matches the spam prevention mechanisms used on blockchains such as Ethereum and Solana.

To map such a game to our mean field model, we approximate the rank (order statistics) by quantile deadlines.
Let $\tau_m := \bar F^{-1}(m/N)$ be the population $m/N$ quantile of completion times under mean field.

We can think of $\bar F$ as the CDF for the rank (order statistic) of a representative competitor.
We define the stepwise payoff $U(X)=r_m$ if $\tau_{m-1} < X \le \tau_m$ for $m\le k$ (with $\tau_0:=0$), and $U(X)=-g$ if $X>\tau_k$.
Then, for a searcher using a service with CDF $F_X(\tau;\lambda)$,
\begin{align*}
\mathbb{E}[U(X)]
&= \sum_{m=1}^k r_m P(\{X \in [\tau_{m-1}, \tau_m]\}) + (-g)P(\{X > \tau_k\}) \\ 
&= \sum_{m=1}^k r_m F_X(\tau_m) - F_X(\tau_{m-1}) -g (1-F_X(\tau_k)) \\
&= -g + (r_k+g)F_X(\tau_k;\lambda) \\ 
&+\sum_{m=1}^{k-1}(r_m-r_{m+1})F_X(\tau_m;\lambda).
\end{align*}
A rational participation constraint is an analogue of~\eqref{eq:Utility}: $p\lambda \le \mathbb{E}[U(X)]$ (and $\mathbb{E}[U(X)]\ge 0$).

We now illustrate the Top-k race by considering a searcher competing with $N = 20$ other searchers for an opportunity that rewards the first $s = 7$ participants with a linearly decaying reward $r_m = r(s - m + 1)/s$ for $m \leq s$, and imposes a penalty of $-g = r/4$ on any searcher who chooses to participate but whose bundle is not included.

Further, we introduce the parameter $\alpha$, which denotes the fraction of competing searchers that, from the perspective of a given searcher, make use of the fast lane. Under a mean-field approximation, the empirical CDF $\bar{F}(t)$ is then modeled as a mixture of the uncoded and rateless-coded CDFs:

\begin{equation*}
    \bar{F}(\tau) = \alpha F_{X_{\text{rateless}}}(\tau) + (1-\alpha) F_{X_{\text{uncoded}}}(\tau)
\end{equation*}

We now illustrate the Top-$k$ race through a representative MEV searcher competition scenario in which a searcher competes with \(N=20\) other searchers for an opportunity that rewards the first \(s=7\) participants. The payoff is assumed to decay linearly with rank, \(r_m = r(s-m+1)/s\) for \(m \le s\), while any searcher whose bundle is not included incurs a penalty of \(-g = r/4\). The parameter choice captures a setting with a moderately large number of competing searchers and a small number of ultimately successful participants, consistent with observed Ethereum order-flow competition and bundle merging behavior~\cite{flashbots2024orderflow,flashbots2021profitableblock}.

Thus, for any moderate number of competing searchers using the fast lane, only the fast lane yields positive expected utility. Moreover, when very few searchers use the fast lane (small $\alpha$), the expected reward from using it remains substantial; if a searcher is effectively the only fast-lane participant, $E[U(X)]$ approaches the maximum attainable reward $r$.

\begin{figure}
    \centering
    \includegraphics[width=\linewidth]{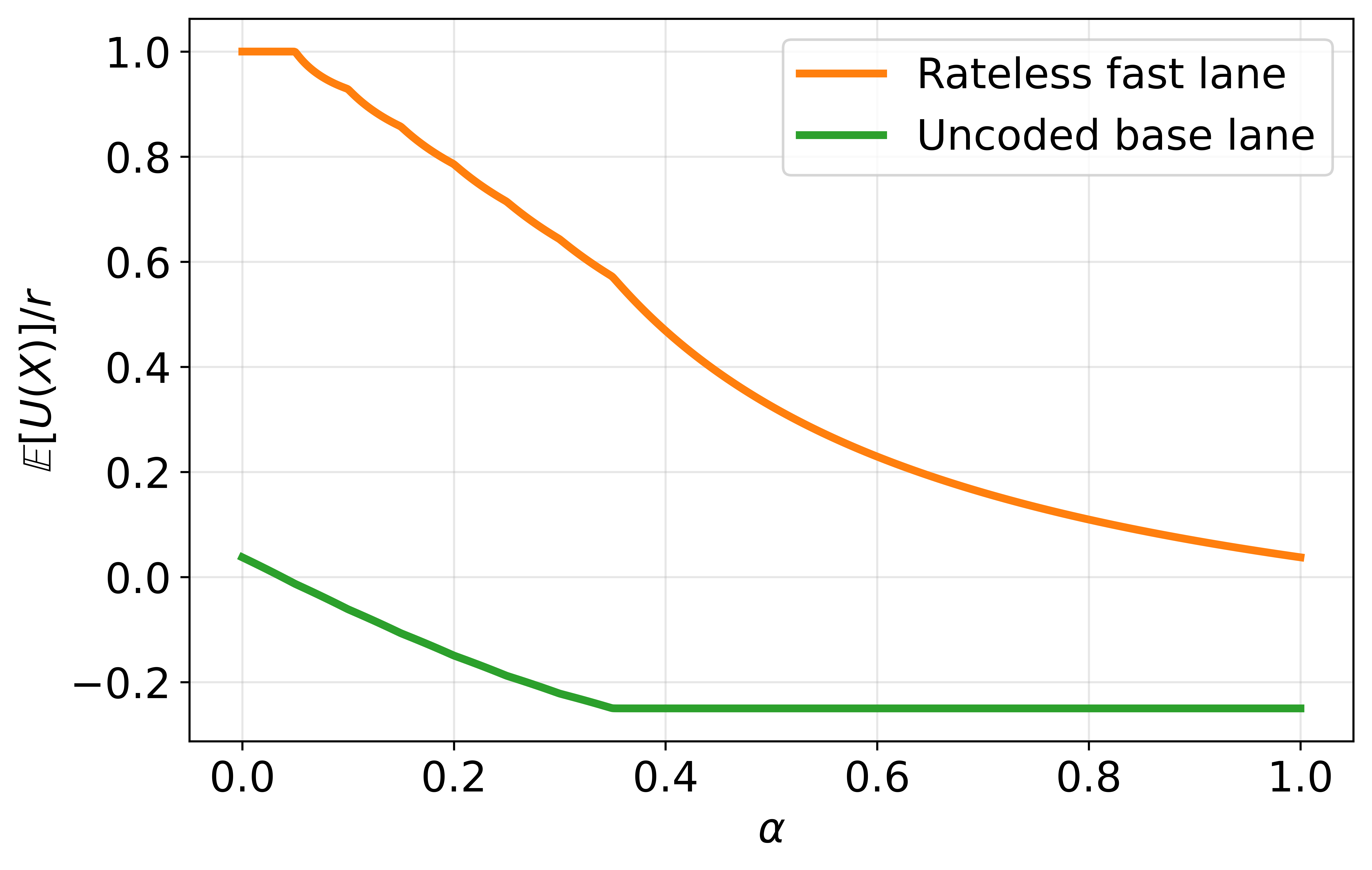}
    \caption{Expected utility in an exemplary Top-k race for a searcher competing with 20 competitors, of which a fraction $\alpha$ use the fast lane. The expected reward for fast-lane and base-lane usage is shown, normalized by the maximum achievable reward $r$.}
    \label{fig:expected_utility_searcher}
\end{figure}

\section{Conclusion}

We presented a pricing model for low-latency payload delivery when rewards depend on when decoding completes. The model makes the decoding-time distribution the central economic object: it determines the probability of meeting deadlines and therefore a rational participant's willingness to pay for additional delivery rate. Within a mean-field framework, we derived explicit price--rate bounds and instantiated them using simple arrival-time models for unsharded transmission and for sharded delivery under three regimes---uncoded transmission, fixed-rate erasure coding, and rateless codes. Further, we exploited the general construction of RLNC to use all types of data shards in addition to a Turbo lane to achieve the best decoding time given all the information available at a node.

The key results are comparative. In our instantiations, rateless RLNC supports higher price bounds than uncoded and fixed-rate schemes because received symbols remain useful for decoding with high probability until reconstruction completes, improving the probability of decoding before a deadline at a given delivered rate. The arrival-time comparison also highlights a qualitative difference between unsharded and sharded delivery: unsharded payloads can have a higher probability of very early completion for small deadlines, while coding across shards reduces variability and improves timely completion at high service levels by reducing the probability of long delays.

We then analyzed a two-lane setting consisting of a base lane and an RLNC fast lane, deriving a fast-lane pricing bound under a fixed base-lane price–rate pair and a maximum fast-lane rate constraint. Across several configurations, we showed that even relatively small additional RLNC rates can increase a node’s utility when the base-lane propagation scheme is sufficiently effective, and we quantified this effect through the corresponding price per shard.

Future work includes refining the arrival model through protocol-specific propagation dynamics, for example via SI-style epidemic dissemination models, and calibrating such models against empirical measurements from real network deployments. It further includes evaluating the utility of the Turbo network in a richer economic framework that captures not only the benefits of accelerated delivery but also the effort and cost incurred by nodes in supplying rate. Further, future work includes extending the competitive model from an exogenous fast-lane usage parameter to an equilibrium framework with heterogeneous participants, endogenous lane choice, and explicit supply dynamics, making it possible to study adoption incentives, equilibrium pricing, and the extent to which the value of accelerated delivery is jointly determined by strategic demand and constrained supply.

\bibliographystyle{IEEEtran}
\bibliography{references}

@book{networkCodingForEngineers,
  title={Network Coding for Engineers},
  author={M{\'e}dard, Muriel and Vasudevan, Vipindev Adat and Pedersen, Morten Videb{\ae}k and Duffy, Ken R},
  year={2025},
  publisher={John Wiley \& Sons}
}

@misc{eip7594,
  author       = {Diederik Loerakker and Proto Lambda and Guillaume Ballet and Dankrad Feist and Vitalik Buterin},
  title        = {EIP-7594: PeerDAS},
  howpublished = {\url{https://eips.ethereum.org/EIPS/eip-7594}},
  year         = {2024},
  note         = {Ethereum Improvement Proposal},
}

@misc{eip4844,
  author       = {Buterin, Vitalik and Feist, Dankrad and Loerakker, Diederik and Kadianakis, George and Garnett, Matt and Taiwo, Mofi and Dietrichs, Ansgar},
  title        = {{EIP-4844: Shard Blob Transactions}},
  howpublished = {\url{https://eips.ethereum.org/EIPS/eip-4844}},
  year         = {2022},
  note         = {Ethereum Improvement Proposal – Proto-Danksharding},
  url          = {https://eips.ethereum.org/EIPS/eip-4844}
}

@article{ho2006random,
  title={A random linear network coding approach to multicast},
  author={Ho, Tracey and M{\'e}dard, Muriel and Koetter, Ralf and Karger, David R and Effros, Michelle and Shi, Jun and Leong, Ben},
  journal={IEEE Transactions on information theory},
  volume={52},
  number={10},
  pages={4413--4430},
  year={2006},
  publisher={IEEE}
}

@article{reed1960polynomial,
  title={Polynomial codes over certain finite fields},
  author={Reed, Irving S and Solomon, Gustave},
  journal={Journal of the society for industrial and applied mathematics},
  volume={8},
  number={2},
  pages={300--304},
  year={1960},
  publisher={SIAM}
}

@article{koetter2003algebraic,
  title={An algebraic approach to network coding},
  author={Koetter, Ralf and M{\'e}dard, Muriel},
  journal={IEEE/ACM transactions on networking},
  volume={11},
  number={5},
  pages={782--795},
  year={2003},
  publisher={IEEE}
}

@article{shokrollahi2006raptor,
  title={Raptor codes},
  author={Shokrollahi, Amin},
  journal={IEEE transactions on information theory},
  volume={52},
  number={6},
  pages={2551--2567},
  year={2006},
  publisher={IEEE}
}

@inproceedings{neu2019babel,
  title={Babel Storage: Uncoordinated Content Delivery from Multiple Coded Storage Systems},
  author={Neu, Joachim and Medard, Muriel},
  booktitle={2019 IEEE Global Communications Conference (GLOBECOM)},
  pages={1--6},
  year={2019},
  organization={IEEE}
}

@inproceedings{hellge2016multi,
  title={Multi-code distributed storage},
  author={Hellge, Cornelius and Medard, Muriel},
  booktitle={2016 IEEE 9th International Conference on Cloud Computing (CLOUD)},
  pages={839--842},
  year={2016},
  organization={IEEE}
}

@article{yang2024buildermarket,
  title={Decentralization of Ethereum's Builder Market},
  author={Yang, Sen and Nayak, Kartik and Zhang, Fan},
  journal={arXiv preprint arXiv:2405.01329},
  year={2024}
}

@inproceedings{schwarzschilling2023timing,
  title={Time Is Money: Strategic Timing Games in Proof-Of-Stake Protocols},
  author={Schwarz-Schilling, Caspar and Saleh, Fahad and Thiery, Thomas and Pan, David and Shah, Nisheeth K. and Monnot, Barnab{\'e}},
  booktitle={Advances in Financial Technologies (AFT)},
  year={2023}
}

@article{oz2023time,
  title={Time Moves Faster When There is Nothing You Anticipate: The Role of Time in MEV Rewards},
  author={{\"O}z, Burak and Kraner, Benjamin and Vallarano, Nicol{\`o} and Kruger, Bingle Stegmann and Matthes, Florian and Tessone, Claudio Juan},
  journal={arXiv preprint arXiv:2307.05814},
  year={2023}
}

@article{pai2023integrated,
  title={Structural Advantages for Integrated Builders in MEV-Boost},
  author={Pai, Mallesh and Resnick, Max},
  journal={arXiv preprint arXiv:2311.09083},
  year={2023}
}

@article{soltani2024eip4844delay,
  title={Delay Analysis of EIP-4844},
  author={Soltani, Pourya and Ashtiani, Farid},
  journal={arXiv preprint arXiv:2409.11043},
  year={2024}
}

@misc{flashbots2025infrastructurerace,
  title={{The Block Auction Infrastructure Race}},
  author={{Flashbots Collective}},
  year={2025},
  note={Flashbots Collective post}
}

@article{wahrstatter2023time,
  title={Time to bribe: Measuring block construction market},
  author={Wahrst{\"a}tter, Anton and Zhou, Liyi and Qin, Kaihua and Svetinovic, Davor and Gervais, Arthur},
  journal={arXiv preprint arXiv:2305.16468},
  year={2023}
}

@article{daian2019flash,
  title={Flash boys 2.0: Frontrunning, transaction reordering, and consensus instability in decentralized exchanges},
  author={Daian, Philip and Goldfeder, Steven and Kell, Tyler and Li, Yunqi and Zhao, Xueyuan and Bentov, Iddo and Breidenbach, Lorenz and Juels, Ari},
  journal={arXiv preprint arXiv:1904.05234},
  year={2019}
}

@online{neroeth2024deep_diving_attestations,
  author  = {Anton Wahrst{\"a}tter},
  title   = {{Deep Diving Attestations: A Quantitative Analysis}},
  date    = {2024-07-09},
  url     = {https://ethresear.ch/t/deep-diving-attestations-a-quantitative-analysis/20020},
  note    = {Ethereum Research post},
}

@online{neroeth2024attestations_block_propagation_timing_games,
  author  = {Anton Wahrst{\"a}tter},
  title   = {{On Attestations, Block Propagation, and Timing Games}},
  date    = {2024-08-14},
  url     = {https://ethresear.ch/t/on-attestations-block-propagation-and-timing-games/20272},
  note    = {Ethereum Research post},
}

@misc{kniep2025solana,
  title={Solana alpenglow consensus},
  author={Kniep, Quentin and Sliwinski, Jakub and Wattenhofer, Roger},
  year={2025}
}

@online{flashbots2024orderflow,
  author       = {{Flashbots}},
  title        = {Illuminating Ethereum's Order Flow Landscape},
  year         = {2024},
  month        = jan,
  day          = {17},
  url          = {https://writings.flashbots.net/illuminate-the-order-flow},
  note         = {Reports an example in which a single transaction attracted 18 competing searcher bids}
}

@online{flashbots2021profitableblock,
  author       = {{Flashbots}},
  title        = {Why Building the Most Profitable Block is Important},
  year         = {2021},
  month        = dec,
  day          = {11},
  url          = {https://writings.flashbots.net/on-the-most-profitable-block},
  note         = {States that the median miner merges a maximum of 3 bundles per block}
}

@book{kleinrock1975queueing,
  author    = {Leonard Kleinrock},
  title     = {Queueing Systems. Volume 1: Theory},
  year      = {1975},
  publisher = {John Wiley \& Sons},
  isbn      = {0471491101}
}

@book{bertsekas1992data,
  author    = {Dimitri P. Bertsekas and Robert G. Gallager},
  title     = {Data Networks},
  edition   = {2},
  year      = {1992},
  publisher = {Prentice Hall},
  isbn      = {0132009161}
}

\end{document}